\documentclass[aps,prd,showpacs,preprintnumbers,twocolumn]{revtex4}
\usepackage{graphicx}
\usepackage{epsfig}
\usepackage{amsmath}
\usepackage{subfigure}
\newbox\pippobox
\def\be{\begin{equation}}
\def\ee{\end{equation}}
\def\bea{\begin{eqnarray}}
\def\eea{\end{eqnarray}}

\def\ee           {{\rm e}}

\def\lag{\langle}
\def\rag{\rangle}
\newcommand{\beq}{\begin{equation}}
\newcommand{\eeq}{\end{equation}}
\newcommand{\beqa}{\begin{eqnarray}}
\newcommand{\eeqa}{\end{eqnarray}}
\newcommand{\beqar}{\begin{eqnarray*}}
\newcommand{\eeqar}{\end{eqnarray*}}

\renewcommand{\eqref}[1]{(\ref{#1})}

\def\c{\chi}

\catcode`\@=12

\begin{document}

%\title{Chiral symmetry breaking and linear confinement in the
%dynamical graviton-dilaton-scalar system with chiral condensate and
%dimension-2 gluon condensate}
\title{A dynamical holographic QCD model for \\
chiral symmetry breaking and linear confinement}
%\title{A dynamical $Dp-Dq$ brane system}
%Regge trajectories and linear quark potential}
%\title{Linear Regge trajectory and linear quark potential in the
%graviton-dilaton-scalar system with chiral condensate and
%dimension-2 gluon condensate}

\author{Danning Li$^{1}$}
%\email{lidn@mail.ihep.ac.cn}
\author{Mei Huang$^{1,2}$}
%\email{huangm@mail.ihep.ac.cn}
\author{Qi-Shu Yan$^{3}$}
%\email{yanqishu@gucas.ac.cn}
\affiliation{$^{1}$ Institute of
High Energy Physics, Chinese Academy of Sciences, Beijing, China}
\affiliation{$^{2}$ Theoretical Physics Center for Science
Facilities, Chinese Academy of Sciences, Beijing, China}
\affiliation{$^{3}$ College of Physical Sciences, Graduate
University of Chinese Academy of Sciences, Beijing, China}
\date{\today }
\bigskip
\date{\today}

\begin{abstract}
A self-consistent holographic QCD model is proposed which can realize both chiral symmetry breaking and confinement, two most important phenomena of QCD. It is pointed out that the model can accommodate both Regge spectra of hadrons and the linear potential between quarks. The model is formulated in the framework of graviton-dilaton-scalar system, where the dilaton field is dual to the dimension-2 gluon condensate and can lead to the linear confinement, while the scalar field corresponds to the quark anti-quark condensate and can explain the property of chiral dynamics.
\end{abstract}


\maketitle

Quantum Chromodynamics (QCD) has been
accepted as the basic theory of describing strong interaction for
more than 30 years. However, it is still a challenge to solve QCD
in non-perturbative region where gauge interaction is strong.
The discovery of the anti-de Sitter/conformal field theory (AdS/CFT)
correspondence and the conjecture of the gravity/gauge duality
\cite{dual} provides a revolutionary method to tackle the problem of
strongly coupled gauge theories. A string description of realistic
QCD has not been successfully formulated yet. Many efforts have been invested in searching for such a realistic description by using the "bottom-up" approach. Recent reviews on 5D holographic QCD models can be found in \cite{review}.

It is well-known that the QCD vacuum is characterized by spontaneous chiral symmetry breaking and color charge confinement. The spontaneous chiral symmetry breaking is
well understood by the dimension-3 quark condensate
$\lag\bar{q}q\rag$ \cite{NJL} in the vacuum, in spite of that, the understanding to confinement remains a challenge. Confinement can be reflected by the Regge trajectories of hadrons \cite{Regge}, which suggests that the color charge can form the string-like structure inside hadrons. It can also be shown by the linear potential between two quarks (either light or heavy) at large distances, {\it i.e.} $V_{\bar{Q}Q}(R)=\sigma_s R$ with $\sigma_s$ the string tension.

A successful holographic QCD model should describe both the Regge trajectories of hadron spectra and linear quark potential. Nonetheless, the models on market can not accommodate both.

Currently a working framework used to describe the Regge
trajectories of hadron spectra is the soft-wall AdS$_5$ model or
Karch-Katz-Son-Stephanov (KKSS) model \cite{Karch:2006pv}, in which the metric is still AdS$_5$, but a quadratic dilaton background is introduced in the 5D action. Its extended versions, by including a
correction in the 5D action \cite{Gherghetta-Kapusta-Kelley,YLWu} or
a modified AdS/CFT dictionary \cite{modified-dc}, are widely used to
investigate the meson spectra and baryon spectra. However, with
AdS$_5$ metric, only Coulomb potential between the two quarks can be produced \cite{Maldacena:1998im}.

The working holographic QCD model to realize the linear quark
potential was proposed by Andreev and Zakharov, who introduced a
positive quadratic correction in the deformed warp factor of ${\rm
AdS}_5$ geometry \cite{Andreev:2006ct}. (The linear heavy quark
potential can also be obtained by introducing other deformed warp
factors as in Refs. \cite{Pirner:2009gr,He:2010ye}.) The positive quadratic correction in the deformed warp factor in some sense behaves as a
negative dilaton background in the 5D action, which motivates the
proposal of the negative dilaton soft-wall model
\cite{Zuo:2009dz,deTeramond:2009xk}. More discussions
on the sign of the dilaton correction can be found in
\cite{KKSS-2,Schmidt-pn-dilaton}.

It is observed that a quadratic background correction in
the metric and dilaton background can lead to the linear quark anti-quark potential and the linear Regge behavior, respectively. It is interesting to explore how to generate both the linear Regge behavior of hadron spectra and linear quark potential in a self-consistent model.
In this work, a holographic QCD model is proposed and formulated in the graviton-dilaton-scalar coupled system.

Let's first briefly describe the graviton-dilaton coupled system, where the dilaton background is expected to be dual to the effective degree freedom in the pure gluon system.
In general, a dilaton background will deform the warp factor of the
metric structure. As described in Ref. \cite{GKN}, the metric structure
of the holographic QCD model and its corresponding dilaton background can be self-consistently
solved from the Einstein equation in the graviton-dilaton coupling system.
It is observed in Ref.\cite{Li:2011hp}, that the positive quadratic correction in
the metric background can describe the pure gluon QCD system quite well. The 5D action of the graviton-dilaton system is defined as
\begin{equation}\label{action-GD}
S_{GD}= \frac{1}{16\pi G_5}\int
 d^5x\sqrt{g_s}e^{-2\phi}\big(R+4\partial_m\phi\partial^m\phi-V_{\phi}\big).
\end{equation}
Where $G_5$ is the 5D Newton constant, $g_s$, $\phi$ and $V_\phi$ are the 5D
metric, the dilaton field and dilaton potential in the string frame, respectively.
Under the quadratic dilaton background
\begin{equation}
\phi=\mu^2 z^2,
\label{dilaton}
\end{equation}
the analytic solution of the dilaton potential in the Einstein frame $V^E_\Phi
=e^{4\phi/3}V_{\phi}$ with $\phi=\sqrt{\frac{3}{8}}\Phi$ takes the form of
\begin{equation}
V_{\Phi}^E=-12~\frac{ _0F_1(1/4;\frac{\Phi^2}{24})^2}{L^2}
+2~ \frac{ _0F_1(5/4;\frac{\Phi^2}{24})^2\Phi^2}{L^2},
\end{equation}
here $L$ the radius of AdS$_5$ and $_0F_1(a;z)$ the hypergeometric function.
In the ultraviolet limit,
$V^E_{\Phi}\overset{\Phi\rightarrow0}{\longrightarrow}-\frac{12}{L^2}-\frac{1}{2}M^2_{\Phi}\Phi^2$
with the 5D mass for the dilaton field $M^2_{\Phi}L^2=4$. From the
AdS/CFT dictionary $\Delta(4-\Delta)=M^2_{\Phi}L^2$, one can derive
its dimension $\Delta=2$. The most likely dimension-2 operator
candidate is $A_{\mu}^2$. It has been pointed out in
Refs.\cite{GC-D2, Xu:2011ud} that the dimension-2 gluon condensate
plays essential role for the linear confinement. Therefore, we can
regard the quadratic dilaton field is dual to the dimension-2 gluon
condensate, i.e., $<A_{\mu}^2>=\mu^2$, and the graviton-dilaton system
describes the pure gluodynamics.

We now add the probe of flavor dynamics on the pure gluodynamic background,
and extend the graviton-dilaton system to the framework of graviton-dilaton-scalar
coupling system, where the scalar field captures chiral dynamics.
The graviton-dilaton-scalar system can be described by the following 5D action:
\begin{equation}
S=S_{GD}+S_{M},
\label{totalaction}
\end{equation}
with $S_{GD}$ given in Eq.(\ref{action-GD}) and $S_{M}$ the KKSS
action for mesons as in \cite{Karch:2006pv} taking the form of
\begin{eqnarray}\label{action1st}
S_{M} &=& -\frac{N_f}{N_c} \int d^5x
 \sqrt{g_s} e^{-\phi} Tr\Big(|DX|^2+V_{X} \nonumber \\
 & &  +\frac{1}{4g_5^2}(F_L^2+F_R^2)\Big).
\end{eqnarray}
Where $X$ and $V_{X}$ are the scalar field and its corresponding potential.
$g_5^2$ is fixed as $12\pi^2N_f/N_c^2$ \cite{Karch:2006pv} and we take
$N_f=2,N_c=3$ in this paper.

We assume the vacuum background is induced by the dilaton field of dimension-2 gluon
condensate and the scalar field of the quark antiquark condensate
$<X>=\frac{\c(z)}{2}$ \cite{Karch:2006pv}, then the vacuum background part of the
action Eq.(\ref{totalaction}) takes the following form
\begin{eqnarray}\label{actionbgrs}
S_{vac}&=&\frac{1}{16\pi G_5}\int d^5x\sqrt{g_s}\big\{
e^{-2\phi}\big(R_s+4\partial_m\phi\partial^m\phi-V_{\phi}\big)\nonumber\\
&-&\lambda e^{-\phi}\big(
\frac{1}{2}\partial_m\c\partial^m\c+V_{\c} \big)\big\},
\end{eqnarray}
with $\lambda=\frac{16\pi G_5 N_f}{N_c L^3}$ and the metric in the
string frame
\begin{equation}
dS_s^2= B_s^2(-dt^2+d\overset{\rightarrow}{x}^2+dz^2),
\end{equation}
where
\begin{equation}
B_s^2\equiv e^{2A_s}\equiv L^2 b_s^2.
\end{equation}

It is easy to derive the following three coupled field equations:
\begin{eqnarray}\label{eom-aphikai}
 -A_s^{''}+A_s^{'2}+\frac{2}{3}\phi^{''}-
 \frac{4}{3}A_s^{'}\phi^{'}-\frac{\lambda}{6}e^{\phi}\c^{'2}&=&0, \\
\phi^{''}+(3A_s^{'}-2\phi^{'})\phi^{'}-\frac{3\lambda}{16}e^{\phi}\chi^{'2} & & \nonumber \\
 -\frac{3}{8}e^{2A_s-\frac{4}{3}\phi}\partial_{\phi}\left(e^{4/3\phi}V_{\phi}
 +\lambda e^{7/3\phi}V_{\c}\right)&=&0, \\
 \label{eom-phi}
 \c^{''}+(3A_s^{'}-\phi^{'})\c^{'}-e^{2A_s}\partial_{\c}V_{\c}&=&0.
 \label{eom-kai}
\end{eqnarray}
If we know the dynamical information of the dilaton field $\phi$ and the scalar field $\chi$,
then the metric $A_s$, the dilaton potential $V_{\phi}$ and the scalar potential $V_{\c}$
should be self-consistently solved from the above three coupled equations.
It is noticed that the graviton-dilaton-scalar system is different from the graviton-dilaton-tachyon system \cite{GDT}, where the metric remains as AdS$_5$.

{\bf\it The UV asymptotic form of $\c(z)$.---}
As proposed in the KKSS model, at ultraviolet(UV), the scalar field takes the
following asymptotic form,
\begin{eqnarray}
\chi(z) \stackrel{z \rightarrow 0}{\longrightarrow} m_q \zeta z+\frac{\sigma}{\zeta} z^3,
\label{chi-IR}
\end{eqnarray}
where $m_q$ is the current quark mass, and $\sigma$ is the quark
antiquark condensate, and $\zeta$ is a normalization constant and is
fixed as $\zeta^2=\frac{N_c^2}{4\pi^2N_f}$. In this paper, we would
fix $m_q=5 {\rm MeV},\sigma=(228 {\rm MeV})^3$ .

{\bf\it The IR asymptotic form of $\c(z)$ constrained from linear quark potential.---}
The linear behavior of quark-antiquark static potential
in the heavy quark mass limit $m_Q\rightarrow \infty$ can describe
the permanent confinement property of QCD. Following
Ref. \cite{Maldacena:1998im}, one can solve
the renormalized free energy of the $\bar{q}q$ system
under the general metric background $A_s$,
\begin{eqnarray}
& &V_{\bar{q}q}(z_0)=\frac{g_p}{\pi z_0}(\int_{0}^{1}
d\nu(\frac{b_s^2(z_0 \nu)z_0^2}{\sqrt{1-\frac{b_s^4(z_0)}{b_s^4(z_0
\nu)}}}-\frac{1}{\nu^2})-1), ~~~ \label{vqqrn} \\
& & R_{\bar{q}q}(z_0)=2 z_0 \int_{0}^{1}
d\nu\frac{1}{\sqrt{1-\frac{b_s^4(z_0)}{b_s^4(z_0\nu)}}}\frac{b_s^2(z_0)}{b_s^2(z_0\nu)}.
\label{Rqq}
\end{eqnarray}
Here $g_p=\frac{L^2}{\alpha}$ and $\alpha$ the 5D effective string tension.
The integrate kernel in Eq.(\ref{vqqrn}) has a
pole at $\nu=1$, expanding the integral kernel at $\nu=1$ one has
\begin{eqnarray}\label{nu-1}
1-\frac{b_s^4(z_0)}{b_s^4(z_0\nu)} =
\frac{4z_0b_s^{'}(z_0)}{b_s(z_0)}(\nu-1)+o((\nu-1)^2)
\end{eqnarray}
From Eqs.(\ref{vqqrn},\ref{Rqq},\ref{nu-1}), one can observe that at
the point $z_c$ when $b_s^{'}(z_c)\rightarrow 0$, then the integral
is dominated by $\nu=1$ region, one can have
\begin{eqnarray}
\frac{V_{\bar{q}q}(z_0)}{R_{\bar{q}q}(z_0)}\overset{z_0\rightarrow z_c}{\longrightarrow} \frac{g_p}{2\pi} b_s^2(z_c). \label{stringtension}
\end{eqnarray}
Therefore, the necessary condition for the linear part of the
$q-\bar{q}$ potential is that there exists one point $z_c$ or one region,
where $b_s^{'}(z)\rightarrow 0,z\rightarrow z_c$ while $b_s(z)$ keeps
finite. For simplicity, we can take the following constraint on the
metric structure at IR:
\begin{equation}
A_s^{'}(z) \stackrel{z \rightarrow \infty}{\longrightarrow} 0,
A_s(z) \stackrel{z \rightarrow \infty}{\longrightarrow} {\rm Const}.
\label{Metric-IR}
\end{equation}
Under the condition of Eq.(\ref{Metric-IR}),
the equation of motion Eq.(\ref{eom-aphikai}) in the IR takes the
following simple form:
\begin{equation}
\frac{2}{3}\phi^{''}-\frac{\lambda}{6}e^{\phi}\c^{'2}=0,
\label{phi-chi-IR}
\end{equation}
which provides a relation between the chiral condensate and
dimension-2 gluon condensate at IR. The asymptotic form of
$\c(z)$ at IR can be solved as:
\begin{equation}
\chi(z)\stackrel{z \rightarrow \infty}{\longrightarrow} \sqrt{8/\lambda}\mu e^{-\phi/2}.
\label{chi-UV}
\end{equation}

To match the asymptotic forms both at UV and IR in Eqs.(\ref{chi-IR}) and (\ref{chi-UV}),
$\c$ can be parameterized as
\begin{eqnarray}\label{chiz}
\c^{'}(z)=\sqrt{8/\lambda}\mu e^{-\phi/2}(1+c_1 e^{- \phi}+c_2
e^{-2\phi})
\end{eqnarray}
with
$c_1=-2+\frac{5\sqrt{2\lambda}m_q\zeta}{8\mu}+\frac{3\sqrt{2\lambda}\sigma}{4\zeta
\mu^3}$ and $c_2=1-\frac{3\sqrt{2\lambda}m_q\zeta}{8\mu}-\frac{3\sqrt{2\lambda}\sigma}{4\zeta
\mu^3}$.

{\bf\it Regge trajectories of mesons.---}
Under the dilaton background Eq.(\ref{dilaton}) and the scalar profile Eq.(\ref{chiz}), the metric structure $A_s(z)$ or $b_s(z)$ can be solved through Eq.(\ref{eom-aphikai}).
Considering the meson fluctuations under the vacuum background described by $A_s, \phi, \c$ as in \cite{Karch:2006pv}, we can split the fields into background part
and perturbation part. For $X$ we would have a scalar perturbation
$s$ and pseudo-scalar perturbation
$\pi$, i.e., $X=(\frac{\c}{2}+s)e^{\i 2\pi^a t^a}$. For the vector
field $V_\mu$ and axial vector field $A_\mu$ part, due to their
vanishing vacuum expectation value, we would use the same notation to denote
the perturbation fields. The equations of motion for perturbation fields
$s,\pi,V_\mu,A_\mu$ can be easily derived. For example, the schrodinger
like equation for vector is given below:
\begin{eqnarray}
& &-v_n^{''}+V_v(z)v_n=m_{n,v}^2v_n, \\
\label{scalar-sn}
& & V_v(z)=\frac{A_s^{''}-\phi^{''}}{2}+\frac{(A_s^{'}-\phi^{'})^2}{4} \label{scalar-vsz} .
\end{eqnarray}
It is noticed that at IR, $V_v(z)=-\mu^2 + \mu^4z^2$,
therefore the solution of Eq.(\ref{scalar-sn}) for high excitations is
$M_n^2=4\mu^2 n$.

\begin{figure}[h]
\begin{center}
\epsfxsize=6 cm \epsfysize=6 cm \epsfbox{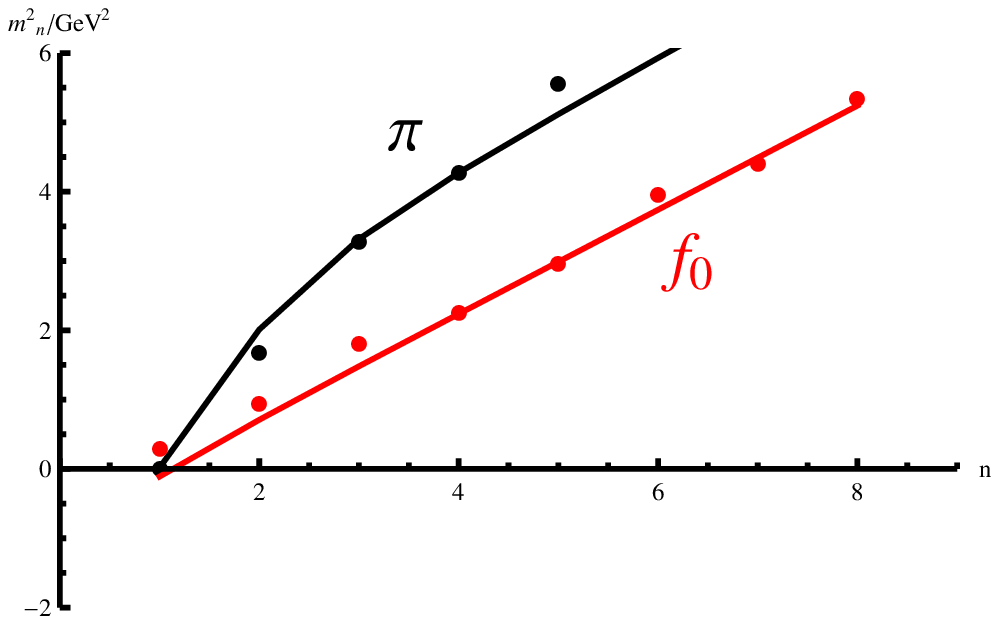}  \\
\epsfxsize=6 cm \epsfysize=6 cm \epsfbox{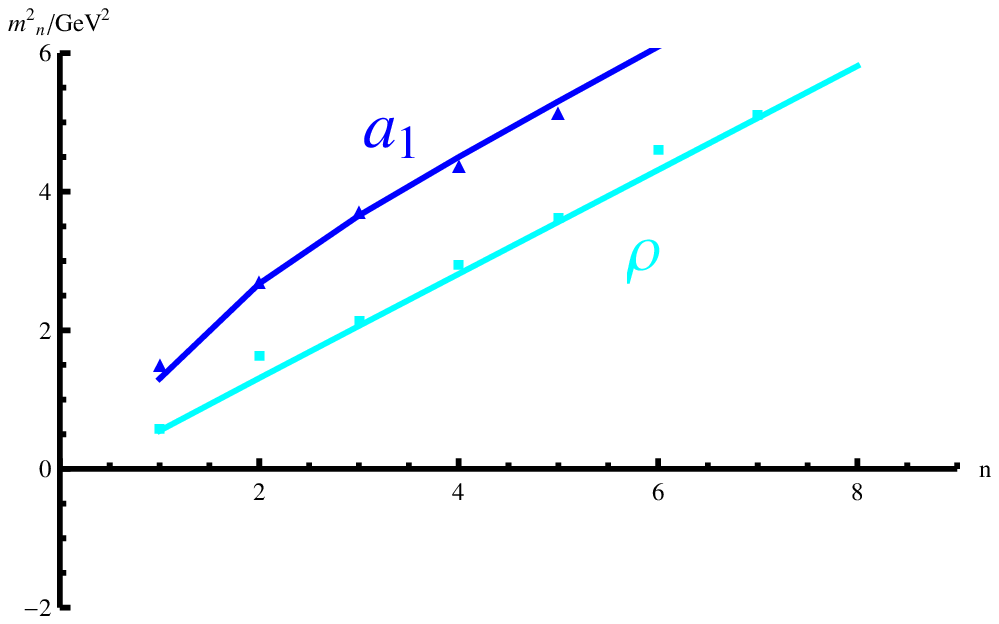}
\end{center}
\caption{A plot of experimental(dot) and model predicted (line) mass
square spectra for the scalar and pseudoscalar mesons $f_0,\pi$ and
vector and axial-vector mesons $\rho, a_1$.} \label{pif0rhoa1-mass}
\end{figure}

By fixing the 5D Newton constant $G_5=\frac{3 L^3}{4}$, one can produce the proper
splitting between the vector and axial vector Regge trajectories.
The produced spectra of scalar $f_0$, pseudoscalar $\pi$, vector $\rho$ and axialvector $a_1$
are shown in Fig.\ref{pif0rhoa1-mass} compared with experimental data \cite{pdg}. It is worth
of mentioning that the experimental data for $f_0$ are chosen as in Ref.
\cite{Gherghetta-Kapusta-Kelley}. With only 4 parameters, all produced meson spectra in the graviton-dilaton-scalar system agree well with experimental data except $f_0(600)$.
It is observed that the Regge trajectories are parallel and the slope is $4\mu^2$
with $\mu=0.43 {\rm GeV}$.

{\bf\it The string tension of the linear quark potential.---}
Under the metric background $A_s(z)$ solved from Eq.(\ref{eom-aphikai}) with the dilaton background Eq.(\ref{dilaton}) and the scalar profile Eq.(\ref{chiz}), the quark potential
can be solved from Eqs.(\ref{vqqrn}) and (\ref{Rqq}). In the UV limit, one can derive the Coulomb potential $V_{\bar{q}q}=-\frac{0.23 g_p}{R_{\bar{q}q}}$ as given in Ref.\cite{Maldacena:1998im}. In the IR limit, we can get the linear potential $V_{\bar{q}q}=\frac{g_p}{2\pi} b_s^2(z_c) R_{\bar{q}q}$. From the solutions in Eq.(\ref{eom-aphikai}), we have $b_s^2 \approx 4 \mu^2$, which indicates that the string tension of the linear quark potential $\sigma_s \sim 4 \mu^2$.

\begin{figure}[h]
\begin{center}
\epsfxsize=6 cm \epsfysize=6 cm \epsfbox{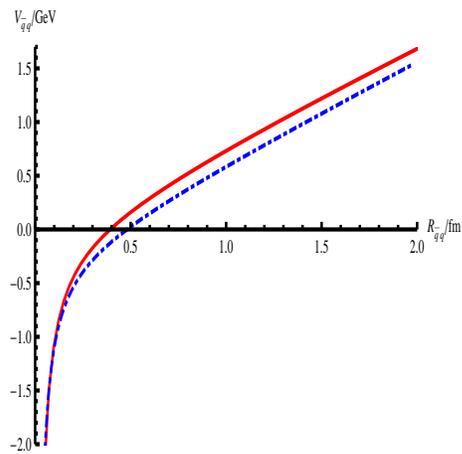}
\end{center}
\caption{$V_{\bar{q}q}$ as a function of $R_{\bar{q}q}$ from our
model (solid line) with $g_p=2.3$ and $\mu=0.43{\rm GeV}$ compared
with the Cornell potential (dot-dashed line). } \label{Vqq}
\end{figure}

The numerical result for the quark potential $V_{\bar{q}q}$ as a function
of quark anti-quark distance $R_{qq}$ is shown by the solid line in
Fig.\ref{Vqq}. The two parameters in Eq.(\ref{vqqrn}) are fixed as
$g_p=2.3$ and $\mu=0.43 {\rm GeV}$. The result agrees with the
Cornell potential (dot-dashed line) \cite{Cornell}
$V^c(R)=-\frac{\kappa}{R}+\sigma_{str}R+V_0$  with $\kappa\approx
0.48$, $\sigma_{str}\approx 0.183 {\rm GeV}^{2}$ and $V_0=-0.25 {\rm
GeV}$.

The last but not the least, we discuss the sign of the dilaton background.
If we choose a negative dilaton background $\phi=-\mu^2 z^2$ as in Ref.\cite{Zuo:2009dz,deTeramond:2009xk}, in the IR limit,
the last term in Eq.(\ref{eom-aphikai}) decreases exponentially to zero, and one can get the asymptotic solution of $A_s(z) \overset{z\rightarrow\infty}{\longrightarrow} -\frac{4}{3}\mu^2z^2+1/2\log(z)$ and
$b_s(z) \sim \sqrt{z} e^{-\frac{4}{3}\mu^2 z^2}\overset{z\rightarrow\infty}{\longrightarrow}0$.
From Eq.(\ref{stringtension}), it's not possible to produce the linear potential with a negative dilaton background. Therefore, one can safely exclude the negative dilaton background in the graviton-dilaton-scalar system.

In summary, we propose a dynamical holographic QCD model, which takes into account  the back-reaction of flavor dynamics on the pure gluodynamic background. To our knowledge, this is the first dynamical holographic QCD model which can
produce both the linear Regge trajectories of hadron
spectra and quark anti-quark linear potential. It is observed that
both the slope of the Regge trajectories and the string tension of
the linear quark anti-quark potential are proportional to the dimension-2
gluon condensate. This result indicates that the linear confinement is
dynamically induced by the dimension-2 gluon condensate. The holographic QCD model offers us a new viewpoint on the relation between the chiral symmetry breaking and
confinement. It is found that the balance between the chiral
condensate and dimension-2 gluon condensate is essential to produce
the correct Regge behavior of hadron spectra. As a byproduct, it is
found that the negative dilaton background can be safely excluded in the
framework of graviton-dilaton-scalar system.

{\bf\it Acknowledgement.---}
We thank valuable discussions with S. He, C. Liu, Y.X. Liu, J.J. Wu, J.B. Wu and
F.K. Xu. This work is supported by the NSFC under Grant
No. 11175251, CAS fellowship for young foreign scientists under
Grant No. 2011Y2JB05, CAS key project KJCX2-EW-N01, K.C.Wong
Education Foundation, and Youth Innovation Promotion Association of CAS.


\begin{thebibliography}{99}

\bibitem{dual} J.~M.~Maldacena,
%``The large N limit of superconformal field theories and supergravity,''
Adv.\ Theor.\ Math.\ Phys.\ \textbf{2}, 231 (1998) [Int.\ J.\
Theor.\ Phys.\ \textbf{38}, 1113 (1999)]; S.~S.~Gubser,
I.~R.~Klebanov and A.~M.~Polyakov,
%``Gauge theory correlators from non-critical string theory,''
Phys.\ Lett.\ B \textbf{428}, 105 (1998); %Lectures on AdS
E. Witten,
%``The large N limit of superconformal field theories and supergravity,''
Adv.Theor.Math.Phys. 2 (1998) 253-291.

\bibitem{review}
  G.~F.~de Teramond and S.~J.~Brodsky,
  %``Hadronic Form Factor Models and Spectroscopy Within the Gauge/Gravity Correspondence,''
  arXiv:1203.4025 [hep-ph].  %%CITATION = ARXIV:1203.4025;%%
Y.~Kim, I.~J.~Shin and T.~Tsukioka,
  %``Holographic QCD: Past, Present, and Future,''
arXiv:1205.4852 [hep-ph].  %%CITATION = ARXIV:1205.4852;%%
A.~Adams, L.~D.~Carr, T.~Schaefer, P.~Steinberg and J.~E.~Thomas,
  %``Strongly Correlated Quantum Fluids: Ultracold Quantum Gases, Quantum Chromodynamic Plasmas, and Holographic Duality,''
arXiv:1205.5180 [hep-th].  %%CITATION = ARXIV:1205.5180;%%

\bibitem{NJL}
  Y.~Nambu,
  %``Quasiparticles and Gauge Invariance in the Theory of Superconductivity,''
  Phys.\ Rev.\  {\bf 117}, 648-663 (1960).

\bibitem{Regge} G.~Veneziano,
%``Construction Of A Crossing - Symmetric, Regge Behaved Amplitude For
%Linearly Rising Trajectories,''
Nuovo Cim.\ A \textbf{57}, 190 (1968); %%CITATION = NUCIA,A57,190;%%
P.D.B. Collins, \textit{An Introduction to Regge Theory and High
Energy Physics}, Cambridge Univ. Press, Cambridge (1975).

%%%%%%%%%%%%%%%% Soft-wall %%%%%%%

%\cite{Karch:2006pv}
\bibitem{Karch:2006pv}
  A.~Karch, E.~Katz, D.~T.~Son and M.~A.~Stephanov,
 % ``Linear confinement and AdS/QCD,''
 Phys.\ Rev.\ D {\bf 74} (2006) 015005  %[hep-ph/0602229].  %%CITATION = HEP-PH/0602229;%%


%\cite{}
\bibitem{Gherghetta-Kapusta-Kelley}
  T.~Gherghetta, J.~I.~Kapusta and T.~M.~Kelley,
%  ``Chiral symmetry breaking in the soft-wall AdS/QCD model,''
Phys.\ Rev.\ D {\bf 79} (2009) 076003; % [arXiv:0902.1998 [hep-ph]].  %%CITATION = ARXIV:0902.1998;%%
 T.~M.~Kelley, S.~P.~Bartz and J.~I.~Kapusta,
%  ``Pseudoscalar Mass Spectrum in a Soft-Wall Model of AdS/QCD,''
  Phys.\ Rev.\ D {\bf 83} (2011) 016002;
  %[arXiv:1009.3009 [hep-ph]].  %%CITATION = ARXIV:1009.3009;%%
T.~M.~Kelley,
 % ``The Dynamics and Thermodynamics of Soft-Wall AdS/QCD,''
 arXiv:1108.0653 [hep-ph].

\bibitem{YLWu}
  Y.~-Q.~Sui, Y.~-L.~Wu, Z.~-F.~Xie and Y.~-B.~Yang,
  %``Prediction for the Mass Spectra of Resonance Mesons in the Soft-Wall AdS/QCD with a Modified 5D Metric,''
  Phys.\ Rev.\ D {\bf 81} (2010) 014024;  % [arXiv:0909.3887 [hep-ph]].  %%CITATION = ARXIV:0909.3887;%%
Y.~-Q.~Sui, Y.~-L.~Wu and Y.~-B.~Yang,
  %``Predictive AdS/QCD Model for Mass Spectra of Mesons with Three Flavors,''
  Phys.\ Rev.\ D {\bf 83} (2011) 065030.  %[arXiv:1012.3518 [hep-ph]].  %%CITATION = ARXIV:1012.3518;%%

\bibitem{modified-dc}
 S.~J.~Brodsky and G.~F.~de Teramond,
  %``Light front hadron dynamics and AdS / CFT correspondence,''
  Phys.\ Lett.\ B {\bf 582}, 211 (2004);
  %[hep-th/0310227].  %%CITATION = HEP-TH/0310227;%%
T.~Branz, T.~Gutsche, V.~E.~Lyubovitskij, I.~Schmidt and A.~Vega,
  %``Light and heavy mesons in a soft-wall holographic approach,''
  Phys.\ Rev.\ D {\bf 82}, 074022 (2010)
  % [arXiv:1008.0268 [hep-ph]].  %%CITATION = ARXIV:1008.0268;%%

%\cite{Maldacena:1998im}
\bibitem{Maldacena:1998im}
  J.~M.~Maldacena,
 % ``Wilson loops in large N field theories,''
   Phys.\ Rev.\ Lett.\  {\bf 80} (1998) 4859
   %[hep-th/9803002].  %%CITATION = HEP-TH/9803002;%%


%%%%%%%%%%%Heavy quark potential %%%%%%%%%%%%%

\bibitem{Andreev:2006ct}
  O.~Andreev and V.~I.~Zakharov,
 % ``Heavy-quark potentials and AdS/QCD,''
  Phys.\ Rev.\  D {\bf 74}, 025023 (2006).
  %%[arXiv:hep-ph/0604204].
  %%CITATION = PHRVA,D74,025023;%%

\bibitem{Pirner:2009gr}
  H.~J.~Pirner and B.~Galow,
  %``Strong Equivalence of the AdS-Metric and the QCD Running Coupling,''
  Phys.\ Lett.\  B {\bf 679}, 51 (2009)
 % [arXiv:0903.2701 [hep-ph]].
  %%CITATION = PHLTA,B679,51;%%
  %\cite{He:2010ye}

\bibitem{He:2010ye}
  S.~He, M.~Huang and Q.~S.~Yan,
  %``Logarithmic correction in the deformed ${\rm AdS}_5$ model to produce the
 % heavy quark potential and QCD beta function,''
 Phys.\ Rev.\ D {\bf 83}, 045034 (2011).
 % arXiv:1004.1880 [hep-ph].
  %%CITATION = ARXIV:1004.1880;%%

%%{Zuo:2009dz,deTeramond:2009xk}
\bibitem{Zuo:2009dz}
  F.~Zuo,
  %``Improved Soft-Wall model with a negative dilaton,''
Phys.\ Rev.\  D {\bf 82}, 086011 (2010).
% arXiv:0909.4240 [hep-ph].
  %%CITATION = ARXIV:0909.4240;%%

\bibitem{deTeramond:2009xk}
  G.~F.~de Teramond and S.~J.~Brodsky,
%  ``Light-Front Holography and Gauge/Gravity Duality: The Light Meson and
  Baryon Spectra,''
  arXiv:0909.3900 [hep-ph].
  %%CITATION = ARXIV:0909.3900;%%

\bibitem{Schmidt-pn-dilaton}
  T.~Gutsche, V.~E.~Lyubovitskij, I.~Schmidt and A.~Vega,
  %``Dilaton in a soft-wall holographic approach to mesons and baryons,''
  Phys.\ Rev.\ D {\bf 85}, 076003 (2012). % [arXiv:1108.0346 [hep-ph]].
  %%CITATION = ARXIV:1108.0346;%%

\bibitem{KKSS-2}
 A.~Karch, E.~Katz, D.~T.~Son and M.~A.~Stephanov,
  %``On the sign of the dilaton in the soft wall models,''
  JHEP {\bf 1104}, 066 (2011).
  % arXiv:1012.4813 [hep-ph].
  %%CITATION = ARXIV:1012.4813;%%

%%%%%%%%%%%%%%
%\cite{Gursoy:2007cb}
\bibitem{GKN}
  U.~Gursoy and E.~Kiritsis,
 % ``Exploring improved holographic theories for QCD: Part I,''
  JHEP {\bf 0802} (2008) 032;
  % [arXiv:0707.1324 [hep-th]].  %%CITATION = ARXIV:0707.1324;%%
  U.~Gursoy, E.~Kiritsis and F.~Nitti,
  %``Exploring improved holographic theories for QCD: Part II,''
   JHEP {\bf 0802} (2008) 019.
   % [arXiv:0707.1349 [hep-th]].  %%CITATION = ARXIV:0707.1349;%%

\bibitem{Li:2011hp}
  D.~Li, S.~He, M.~Huang and Q.~-S.~Yan,
 % ``Thermodynamics of deformed AdS$_5$ model with a positive/negative quadratic correction in graviton-dilaton system,''
 JHEP {\bf 1109} (2011) 041.
 % [arXiv:1103.5389 [hep-th]].
 %%CITATION = ARXIV:1103.5389;%%

\bibitem{GC-D2}
  F.~V.~Gubarev, L.~Stodolsky, V.~I.~Zakharov,
  %``On the significance of the vector potential squared,''
  Phys.\ Rev.\ Lett.\  {\bf 86}, 2220-2222 (2001);
  K.~I.~Kondo,
  %``Vacuum condensate of mass dimension 2 as the origin of mass gap and  quark
  %confinement,''
  Phys.\ Lett.\  B {\bf 514}, 335 (2001).
%  [hep-ph/0010057].

\bibitem{Xu:2011ud}
  F.~Xu and M.~Huang,
  %``Electric and magnetic screenings of gluons in a model with dimension-2 gluon condensate,''
   arXiv:1111.5152 [hep-ph].
   %CITATION = ARXIV:1111.5152;

\bibitem{GDT}
  B.~Batell and T.~Gherghetta,
 % ``Dynamical Soft-Wall AdS/QCD,''
 Phys.\ Rev.\ D {\bf 78} (2008) 026002;
 % [arXiv:0801.4383 [hep-ph]].  %%CITATION = ARXIV:0801.4383;%%
%\bibitem{Kapusta:2010mf}
  J.~I.~Kapusta and T.~Springer,
  %``Potentials for soft wall AdS/QCD,''
  Phys.\ Rev.\ D {\bf 81}, 086009 (2010);
  %[arXiv:1001.4799 [hep-ph]].  %%CITATION = ARXIV:1001.4799;%%
T.~M.~Kelley,
 % %``The Thermodynamics of a 5D Gravity-Dilaton-Tachyon Solution,''
 arXiv:1107.0931 [hep-ph].  %%CITATION = ARXIV:1107.0931;%%

\bibitem{pdg}
  C.~Amsler {\it et al.}  [Particle Data Group Collaboration],
 % ``Review of Particle Physics,''
 Phys.\ Lett.\ B {\bf 667} (2008) 1.
 %%CITATION = PHLTA,B667,1;%%

\bibitem{Cornell}
  E.~Eichten, K.~Gottfried, T.~Kinoshita, K.~D.~Lane and T.~M.~Yan,
  %``Charmonium: Comparison With Experiment,''
  Phys.\ Rev.\  D {\bf 21}, 203 (1980).
  %%CITATION = PHRVA,D21,203;%%


\end{thebibliography}
\end{document}